\documentclass[sigconf]{acmart}

\usepackage{booktabs}
\usepackage{graphicx} 
\graphicspath{{figs/}{figures/}{pictures/}{images/}{./}} 
\usepackage{enumitem}

\title{KinemaFX: A Kinematic-Driven Interactive System for Particle Effect Exploration and Customization}

\author{Yifei Zhang}
\affiliation{%
   \institution{Fudan University}
   \country{China}
   }
\email{maglaris1212@gmail.com}

\author{Lin-Ping Yuan}
\authornote{corresponding author}
\affiliation{%
   \institution{The Hong Kong University of Science and Technology}
   \country{China}
   }
\email{yuanlp@cse.ust.hk}

\author{Yuheng Zhao}
\affiliation{%
   \institution{Fudan University}
   \country{China}
   }
\email{yuhengzhaocn@gmail.com}

\author{Jielin Feng}
\affiliation{%
   \institution{Fudan University}
   \country{China}
   }
\email{jielin0217@gmail.com}

\author{Siming Chen}
\authornotemark[1]
\affiliation{%
   \institution{Fudan University}
   \country{China}
   }
\email{simingchen@fudan.edu.cn}

\begin{abstract}

Particle effects are widely used in games and animation to simulate natural phenomena or stylized visual effects. 
However, creating effect artworks is challenging for non-expert users due to their lack of specialized skills, particularly in finding particle effects with kinematic behaviors that match their intent.
To address these issues, \rv{we propose a conceptual model of particle effects that captures both semantic features and kinematic behaviors. Based on the model,} we present KinemaFX, a kinematic-driven interactive system, to assist non-expert users in constructing customized particle effect artworks.
KinemaFX adopts a workflow powered by Large Language Models (LLMs) that supports intent expression through combined semantic and kinematic inputs, while enabling implicit preference-guided exploration and subsequent creation of customized particle effect artworks based on exploration results. 
Additionally, we developed a kinematic-driven method to facilitate efficient interactive particle effect search within KinemaFX via structured representation and measurement of particle effects.
To evaluate KinemaFX, we illustrate usage scenarios and conduct a user study employing an ablation approach. Evaluation results demonstrate that KinemaFX effectively supports users in efficiently and customarily creating particle effect artworks.

\end{abstract}


\begin{teaserfigure}
  \centering
  \includegraphics[width=\textwidth]{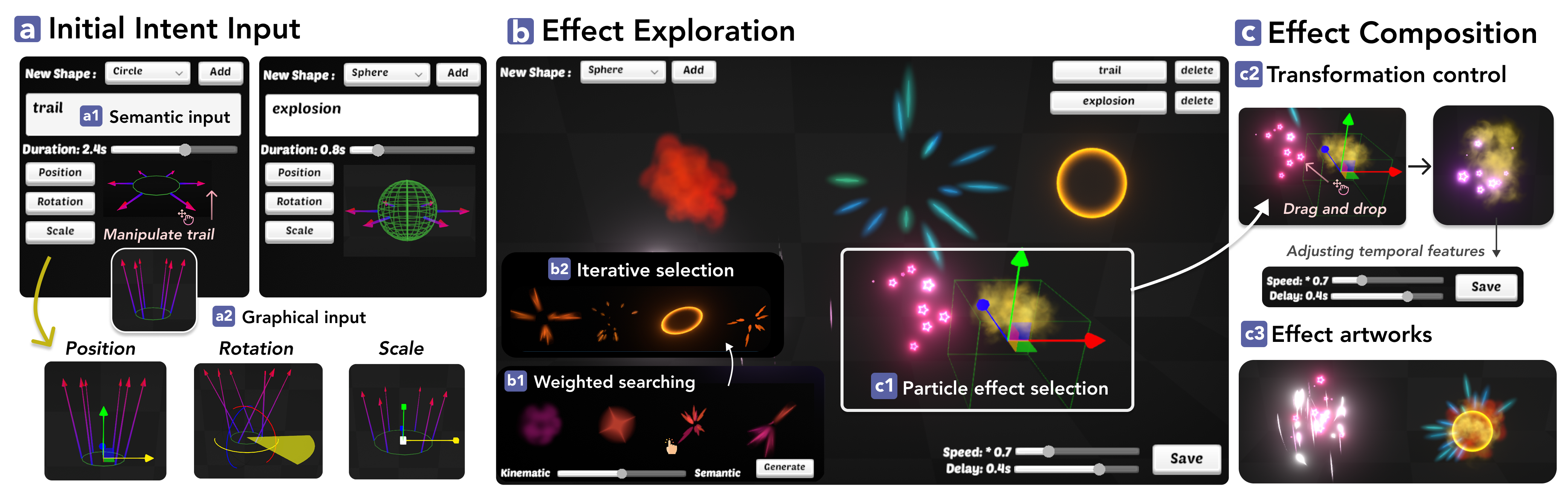}
  \caption{An overview of KinemaFX. KinemaFX supports interactive particle effect exploration and customization through three stages: (a) User Intent Input. Users can express their initial exploration intent by combining semantic input (a1) and graphical input (a2). (b) Effect Exploration. Kinematic supports particle effects searching based on controllable weights of semantic and kinematic similarity (b1). Users iteratively explore the space by selecting satisfying effects, thereby implicitly conveying their preferences(b2). (c) Effect Composition. From the explored particle effects, users can select (c1) and control individual effects' transformations and temporal features (c2) to compose effect artworks (c3).}
  \Description{}
  \label{fig:teaser}
\end{teaserfigure}

\newcommand{\rv}[1]{{\color{black}#1}}

\newcounter{goalcounter}

\newcommand{\Ggoal}[1]{
    \refstepcounter{goalcounter} 
    \textbf{G\thegoalcounter.} #1
}

\newcounter{findingcounter}

\newcommand{\Ffinding}[1]{
    \refstepcounter{findingcounter} 
    \textbf{F\thefindingcounter.} #1
}

\copyrightyear{2025}
\acmYear{2025}
\setcopyright{acmlicensed}
\acmConference[UIST '25]{The 38th Annual ACM Symposium on User Interface Software and Technology}{September 28-October 1, 2025}{Busan, Republic of Korea}
\acmBooktitle{The 38th Annual ACM Symposium on User Interface Software and Technology (UIST '25), September 28-October 1, 2025, Busan, Republic of Korea}
\acmDOI{10.1145/3746059.3747734}
\acmISBN{979-8-4007-2037-6/2025/09}

\begin{document}

\maketitle
\section{Introduction}
Particle effects, a technique that simulates complex visual phenomena through the aggregation of numerous microscopic particles, are extensively utilized in domains such as gaming and cinematography to replicate natural occurrences, including flames, smoke, and explosions~\cite{jeffrey2020ves, finance2015visual}. 
A particle effect artwork typically consists of multiple relatively independent particle effects that need to be individually constructed and then combined.
These particle effects typically possess independent attributes such as position, velocity, lifespan, and color, which are managed and rendered via a particle system~\cite{zhang2017game, hastings2008interactive}.
Currently prevalent particle systems encompass Shuriken~\cite{unityshuriken} and Visual Effect Graph~\cite{unityvfx} within the Unity engine, as well as Cascade~\cite{unrealcascade} and Niagara~\cite{unrealniagara} in Unreal Engine. 
\rv{By embedding powerful particle systems into mainstream game engines, these platforms have lowered the barrier for individual creators and small teams, thereby amplifying the need for intuitive and accessible tools for particle effects creation.}

\rv{Many users who need to create particle effects, such as game and level designers, indie developers, educators, and content creators, lack strong artistic skills or expertise in specialized tools. They require accessible solutions to prototype effects, teach visual concepts, or enhance multimedia content.}
However, the particle systems are tailored for particle effects specialists, leaving general developers lacking the requisite professional skills and aesthetic acumen to directly construct particle effects using the application programming interface of these particle systems. Moreover, using pre-made effect artworks directly often fails to meet developers' customization needs.
For general developers, a more feasible approach involves searching for suitable particle effects to build their customized effect artworks. However, relying on existing search tools to accomplish this is inefficient and inflexible~\cite{chueca2024search}.
Therefore, there is a need for a method that does not rely on developers' specialized background in effects to support them in efficiently and customarily creating particle effect artworks.

Large Language Models (LLMs), owing to their exceptional analytical and creative capabilities after pre-training, are capable of providing users with extensive professional knowledge and supporting flexible interaction modalities, thus being widely employed in design tasks~\cite{zhou2024large}, including image design~\cite{son2024genquery, chen2024autospark}, 3D scene design~\cite{wang2024chat2layout, fu2024anyhome}, and animation design~\cite{qian2024shape, angert2023spellburst}. However, the target users of these studies are domain experts capable of providing high-quality, explicit user inputs, and the design objects do not involve entities with complex kinematic behavior, such as particle effects. Existing human-AI collaborative approaches for dynamic effect creation are predominantly limited to two-dimensional graphics~\cite{angert2023spellburst, tseng2024keyframer}. The limited capability of LLMs in understanding three-dimensional dynamic visual effects constrains the application of these methods to the domain of particle effects. Little is known about how to support non-expert users in efficiently and customarily constructing complex three-dimensional dynamic effects based on LLMs.

In this work, we aim to fill this gap by developing an interactive system based on LLMs to support non-expert users in creating effect artworks that align with their intentions. 
We conducted a formative study comprising three parts. Firstly, we collected a dataset containing 839 particle effects from the Internet. Secondly, to understand the challenges and needs of non-expert users when constructing particle effect artworks, we carried out semi-structured interviews involving six creators. Thirdly, we performed a qualitative analysis of some key characteristics of particle effects.
The findings revealed that explicitly expressing intentions through semantics and navigating the exploration process proves challenging. This poses the first challenge of supporting users to express their intent flexibly and comprehensively while enabling more efficient user-guided exploration. Also, we found that particle effects' kinematic behavior, though crucial, is difficult to represent semantically or measure through LLMs. This presents the second challenge of effectively incorporating kinematic behavior into both the representation and searching of particle effects.

\rv{We propose a conceptual model of particle effects to capture both semantic features and kinematic behaviors.} We then propose KinemaFX (Fig.~\ref{fig:teaser}), an interactive system for effect artwork exploration and customization with a kinematic-driven method to enhance particle effect search. KinemaFX supports flexible integration of both semantic and kinematic inputs, and an exploration process of the particle effect design space guided by users' implicit preferences. To enable efficient and flexible effect search within KinemaFX, we introduce a structured representation and search method for particle effects that combines both semantic and kinematic information.


To address the first challenge, based on feedback from semi-structured interviews and results from qualitative experiments, we designed an intent expression module combining semantic text input with graphical input including simple shapes and arrowed lines. To support flexible user guidance of the exploration process, KinemaFX allows implicit preference expression through intermediate results. With LLM assistance, this iterative exploration alternates between local exploration and directional exploration, enabling efficient alignment with user preferences while avoiding local optima.
To solve the second challenge, we abstract the kinematic behavior of particle effects as dynamic changes of primitive shapes and represent them as mathematical expressions in a spherical coordinate system. We combine this with semantic embeddings to form a structured representation of particle effects. This representation, working in concert with the LLM, enables flexible distance measurement combining both semantic and kinematic information, and alignment of kinematic behaviors between particle effects. Also, the seamless translation of user intent into structured representations allows them to serve as an intermediary that connects user intentions with concrete particle effect instances.

To evaluate KinemaFX, we conducted a within-subjects user study employing an ablation approach with 16 participants. 
The results demonstrated that KinemaFX effectively supports users in expressing intent, exploring diverse effects, and creating satisfying particle effect compositions with reduced mental effort.

The contributions of this work are summarized as follows:

\begin{itemize}[leftmargin=*]
    \item \rv{We propose a conceptual model of particle effects and present KinemaFX based on it.} KinemaFX is an interactive system based on LLMs that enhances particle effect exploration and customization by supporting users' implicit preference expression.
    \item We develop a structured representation and distance metric method for particle effects to support efficient and flexible interactive search and alignment.
    \item We validate that KinemaFX facilitates users in particle effect exploration and customization through a user study and a gallery of created effect artworks.
\end{itemize}
\section{Related Work}

In this section, we summarize the related work about visual effect creation, human-AI collaborative visual design, and facilitating user-guided exploration.

\subsection{Effect Artwork Creation}
Visual effects are indispensable components of modern digital media, film production, and game development. They enhance visual expression and user experience by simulating natural phenomena or creating surreal effects. Game engines such as Unity and Unreal Engine, as well as visual effects software like Houdini and After Effects, provide powerful toolkits and flexible frameworks that enable developers to create high-quality effects. Existing research systematically introduces the principles of constructing visual effects~\cite{raappana2018great, acuna2024visual, mattila2018visual}, and illustrates examples of building visual effects using particle systems~\cite{zhang2017game, vaaraniemi2016example}. However, even though some professional tools such as particle system parametrization with bimanual hand gestures~\cite{sato2022particle} and node-based particle systems~\cite{luque2022node} try to lower technical barriers and learning costs, the physical simulations, mathematical modeling, and artistic design involved in visual effects creation still pose significant challenges for non-expert users.

To address the issue that manually constructing visual effects requires significant effort and artistic expertise, Energy-Brushes~\cite{xing2016energy} supports users in drawing with coarse-scale energy brushes, which serve as control gestures to drive detailed flow particles representing local velocity fields. 
MagicalHands~\cite{arora2019magicalhands} further enables gesture-based control over behaviors of particle effects in three-dimensional virtual space. However, these methods for controlling visual effects only provide a few dimensions of control, making them unsuitable for constructing more diverse visual effects.  
Chueca et al.~\cite{chueca2024search} use encoding to represent the multi-dimensional parameters of particle effects and optimize them through a genetic algorithm that includes a fitness function based on human feedback. 
This approach enables rapid exploration, but lacks effective measurement of particle effects' kinematic behaviors and control over iteration directions during the process.

Instead of treating particle effects as parameter vectors, \rv{our proposed conceptual model} abstracts the kinematic behavior of particle effects as shape transformations in spherical coordinates, enabling computationally efficient measurement that better aligns with human visual perception. We integrate this kinematic representation and metric with LLM-processed semantic information to facilitate more comprehensive and intuitive interactive exploration of particle effects.

\subsection{Human-AI Collaborative Visual Design}
Human-AI collaborative visual design is a novel design paradigm that combines human creativity with artificial intelligence technology, aiming to enhance design efficiency, lower technical barriers, and inspire innovative ideas through human-AI collaboration. This model has been widely applied in the field of visual design~\cite{huang2024plantography, yuan2021infocolorizer, angert2023spellburst}.
\rv{Particle effect design involves intricate kinematic behaviors that demand precise control, posing challenges for conventional approaches like direct manipulation~\cite{shi2025brickify} and semantic input~\cite{angert2023spellburst, chen2024autospark} to be effectively applied in this field. For instance, achieving smooth and accurate spatiotemporal motion of multiple objects within a 3D environment via direct manipulation~\cite{masson2024directgpt} or sketch-based interfaces~\cite{xing2016energy} requires significant user expertise and effort. Conversely, conveying these complex motions through semantic text input often leads to imprecise communication of the detailed kinematic information.}
Specifically, human-AI collaborative particle effect design faces two challenging issues: how users express visual-related intentions and how to align with these intentions.  

For the first issue, Park et al.~\cite{park2024we} point out that since people are visual thinkers, not verbal thinkers, there are limitations in interacting with AI through text input for design. 
To address the limitations of textual input, PromptCharm~\cite{wang2024promptcharm} and DesignPrompt~\cite{peng2024designprompt} facilitate text-to-image creation through multimodal prompt engineering and refinement. 
GenQuery~\cite{son2024genquery} supports vague user intentions by generating images as intermediaries for search. 
PromptPaint~\cite{chung2023promptpaint} allows users to go beyond language to mix prompts that express challenging concepts. However, these methods essentially provide visual examples and construct more effective prompts, lacking the ability to freely control the expression of visual intentions. 
To enable more direct expression of visual intentions, WorldSmith~\cite{dang2023worldsmith} and DirectGPT~\cite{masson2024directgpt} help users express spatial information through direct manipulation of visual elements. 
However, these methods only support the expression intentions of simple planar regions and cannot handle complex intentions such as 3D dynamics. 
\rv{Logomotion~\cite{liu2025logomotion} allows for the 2D animation design of multiple objects using semantic input; however, this design approach describes the motion of each object relatively independently in natural language, making it difficult to handle the design of a large number of objects with complex 3D motions.}
To allow users to express intentions more naturally, we support users in expressing kinematic-related intentions not only through semantic input but also by controlling 3D graphics.  

For the second issue, to align visual design results with user intentions, LayoutPrompter~\cite{lin2023layoutprompter} improves visual layout effects by measuring and using the input constraints most similar to the user's input as examples. 
SceneMotifCoder~\cite{tam2024scenemotifcoder} achieves high-quality spatial layouts by using LLMs to learn the mapping from semantics to spatial relationships. 
However, these works focus on static layout problems and cannot address the complex 3D kinematic behavior involved in particle effects. 
\rv{Our proposed conceptual model} simplifies the complex 3D dynamic motions of particle effects from a visual perspective, thereby supporting intuitive interactive control over the exploration of complex kinematic behaviors.
We leverage LLMs and a proposed kinematic behavior distance metric to support the alignment of results. Therefore, KinemaFX better searches for and controls the kinematic behavior of particle effects in a way that aligns with human visual intentions.

\subsection{Facilitating User-guided Exploration}
Visual design often involves a vast design space, necessitating flexible tools and methods to support users in autonomously exploring this space. Existing exploration methods include controlling category selection~\cite{choi2018stargan}, adjusting continuous semantic scales via sliders~\cite{chiu2020human, chung2023artinter}, and explorable galleries~\cite{zhang2021method, yuan2025personalized}. 
BO as Assistant~\cite{koyama2022bo} supports design space exploration by monitoring user preferences reflected in sliders and applying Bayesian optimization. However, these methods limit options to a constrained set.

The reasoning and generative capabilities of LLMs enable flexible semantic-based exploration of the design space, but also raise the challenge of how to support users in conducting efficient and divergent exploration. 
3DALL-E~\cite{liu20233dall} allows users to construct different prompts by providing keywords across various dimensions. However, this approach is not conducive to rapid, large-scale exploration of the design space.  
To enable more comprehensive exploration, Luminate~\cite{suh2024luminate} and Prompting for Discovery~\cite{almeda2024prompting} extract dimensions from the design space to structure prompts, facilitating efficient dimensional exploration of the LLM output space. However, these works lack consideration of user feedback, making them not suitable for iterative exploration. 
AutoSpark~\cite{chen2024autospark} assists designers in emotional need alignment, design intention expression, and prompt crafting using generative AI and a semantic network. However, users need to actively construct prompts to express their preferences, which limits the direct integration of user intentions into the iterative process.  
RoomDreaming~\cite{wang2024roomdreaming} supports user preference-guided exploration by directly incorporating users' preferred visual solutions into the iterative process. However, users' preferred visual solutions are first converted into text and then reused for text-to-image generation. This re-conversion process introduces bias in integrating user preferences.  
 
Unlike these works, KinemaFX enables users to guide the iterative exploration process through intermediate result selection, allowing preference integration without the cognitive burden of explicitly expressing preference.
Furthermore, to address the bias introduced by converting visual preferences into text, KinemaFX supports the extrapolation of particle effects based on semantics and kinematic behavior in the direction of user preferences.
\section{Formative Study}
We conducted a formative study to better understand the challenges for non-expert users in particle effect exploration and customization. We first collected a dataset of particle effects from online sources. Then, we conducted semi-structured interviews with six creators to understand how non-expert users describe, observe, and explore particle effects, as well as the challenges they face in the process. Based on the feedback from these interviews, we analyzed the key characteristics of particle effects and synthesized our findings to establish the design goals for our methods.

\begin{figure}[tb]
 \centering
 \includegraphics[width=\columnwidth]{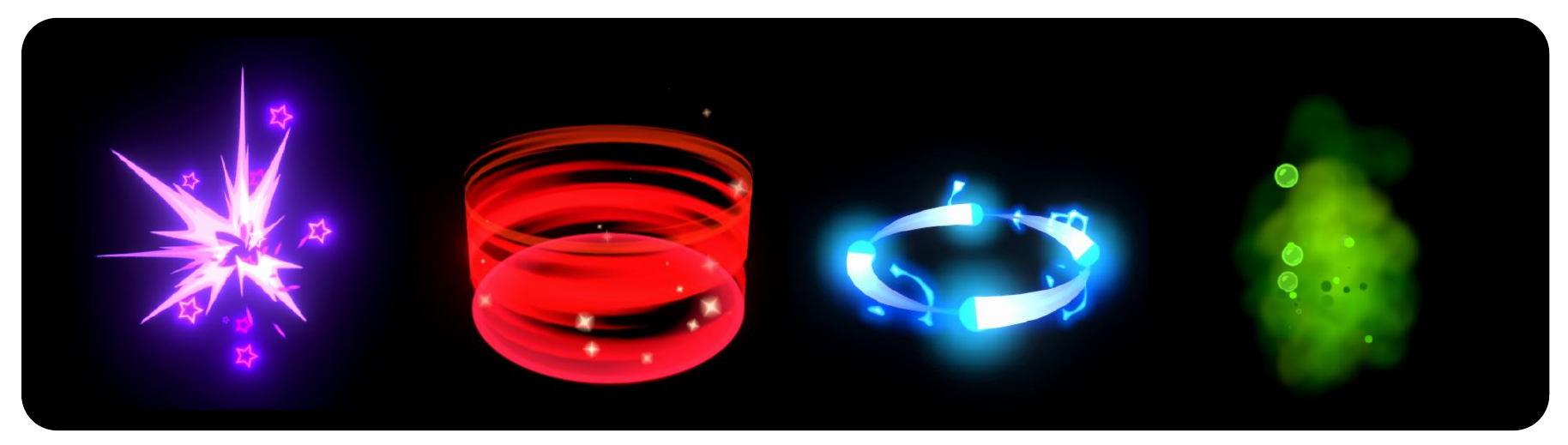}
 \caption{Effect artworks from the collected dataset. Our dataset consists of 147 particle effect artworks, encompassing a total of 839 individual particle effects.}
 \label{fig:examples}
 \Description{Examples of particle effects used in the study.}
\end{figure}

\subsection{Asset Collection}\label{sec:asset}
With the widespread use of particle effects in gaming and animation, numerous design resources are available online. To ensure the dataset's practicality and thematic diversity, we selected the Unity Asset Store as our primary source and curated a dataset from Hovl Studio and Lana Studio collections. The final dataset consists of 147 particle effect artworks spanning 19 themes \rv{(including Generic Area, Backlight, Burst, Fire, Fog, Loot, Orbs, Ranged Attack, Regeneration, States, AoE Effects, Character Auras, Hits and Explosions, Magic Circles, Portals, as well as stylistically distinct Shields and Slash Effects from two different studios)}, encompassing a total of 839 individual particle effects. Our selection criteria were based on:
1) The Unity Asset Store's official review process ensures high-quality assets. The two chosen studios specialize in professional particle effect design.
2) The broad thematic coverage of the selected effects represents commonly used styles and categories in game development.
We showcase several example effect artworks from the collected particle effect assets in Fig.~\ref{fig:examples}.

\subsection{Semi-Structured Interview}

\subsubsection{Method}
We recruited six creators (C1-C6; Age: 19-31; 2 female and 4 male) for our study. Three creators (C1-C3) were researchers in 3D and immersive environment development, while the other three (C4-C6) had amateur game development experience. All creators had some experience with Unity engine, with two creators (C2, C3) possessing basic experience in using particle effects.

The study was conducted through individual interviews, either in-person or via online meetings, with all sessions recorded after obtaining creators' consent. We first inquired about their familiarity with particle effects and their previous creation methods. After introducing our research objectives and methods, we asked the creators to sequentially complete three tasks to explore how users observe and describe particle effects, articulate their intentions for particle effects, and explore particle effects:
1) Task 1: Observation and description of particle effects.
2) Task 2: Conceptualization and description of original particle effects.
3) Task 3: Exploration of particle effect dataset.

For Task 1, we selected 8 effect artworks containing 27 distinct particle effects from the dataset collected in Sec.~\ref{sec:asset}. Creators observed these effects sequentially and described them using their preferred approach. In Task 2, creators were asked to conceptualize an effect artwork and describe each component particle effect. Task 3 provided access to the particle effect dataset with an LLM-based semantic search system, where creators input text descriptions to retrieve matching effects through semantic embedding matching.

Throughout the tasks, we encouraged think-aloud protocols. After each task, creators reflected on their thought processes, discussing their approaches to expressing intentions and any challenges encountered. Additionally, they were asked to identify which characteristics they prioritized when observing and searching for particle effects and how these features could be effectively communicated. We also reviewed the results with them to discuss how retrieved particle effects could be combined into a cohesive effect artwork. The session lasted approximately 45 minutes.

\subsubsection{Findings}

\Ffinding{\textbf{Particle effects' kinematic behavior is crucial but difficult to represent through semantics.}}
In Task 1, all creators (C1-C6) consistently prioritized color attributes, perceived emotional qualities, and general motion patterns when describing particle effects.
The description of perceived effects primarily employed two approaches: hypothetical usage scenarios (C1, C3, C4, C5) and stylistic interpretations (C1, C4, C6). For instance, C4 noted: \textit{``This effect evokes a magical and mystical sensation, which I might employ to highlight special treasure items in a game.''} When characterizing kinematic behavior, creators adopted a hierarchical descriptive approach: first outlining spatial scope and overall shape, then specifying movement direction (C1-C6). Typical descriptions included \textit{``a light ring expanding outward from the center''} (C3) and \textit{``stars rising from the bottom''} (C1). Some creators provided additional trajectory details, such as \textit{``spiraling upward''} (C2) or \textit{``emitting along straight paths''} (C5).
Temporal characteristics were conveyed through concise terminology, including duration (\textit{``instantaneous''} - C1, \textit{``lasting approximately 2-3 seconds''} - C3) and motion sequencing. Creators universally acknowledged the necessity of kinematic descriptions, with one remarking: \textit{``Relying solely on style and emotional impression makes the description too abstract''} (C2). However, upon reflection, creators recognized their tendency to omit kinematic details: \textit{``I typically summarize kinematic behavior with just one or two words rather than describing specifics''} (C6). This abstraction resulted in semantic descriptions that could match multiple visually distinct particle effects.
Notably, creators reported substantial difficulties in verbally articulating kinematic behavior: \textit{``I struggle to describe the dynamic shapes I observe''} (C3). The semantic representation proved particularly inefficient for motion details, as one creator explained: \textit{``It would require lengthy paragraphs to accurately describe the complete motion characteristics''} (C5). These findings collectively demonstrate the inherent limitations of semantic representation for capturing particle effects' kinematic behavior.
\label{F1}

\Ffinding{\textbf{Explicitly expressing intentions is a challenge and graphical input is urgently needed.}}
In Task 2, some creators quickly formed a relatively clear idea for their special effects design (C2, C4, C5), while others only had abstract concepts and fragmented elements (C2, C3, C6). \textit{``I have a clear idea of the color palette and usage scenario in my design, but the finer details remain vague''} (C3). When articulating their design concepts verbally, creators encountered a common difficulty: struggling to find the right words. \textit{``I had to deliberate over many of the terms for my design concept''} (C1); \textit{``I couldn’t think of precise wording—many details are hard to describe with language alone''} (C2). For example, C5 wanted to design a ``convergence'' effect but later found it challenging to accurately convey the starting and ending shapes and scope of this ``convergence'' through words.  
The creators employed non-verbal methods to articulate their design concepts, using hand gestures to depict the shapes and movements of particle effects (C2), and sketching simple diagrams and lines on paper to outline the scope and trajectories of particle effects—including circles (C1, C5), spheres (C4, C6), cylinders (C1, C3), and directional lines (C1, C3, C4, C5, C6). For instance, C1 drew a circle to represent the blast radius of an explosion effect and several rays at an angle to the horizontal plane to indicate the splash direction. Despite having no formal drawing background, he could quickly produce such simple sketches, though he noted, \textit{``I feel it would be much more difficult to describe this idea clearly with words alone.''}
The combination of semantic and graphical input was the most widely endorsed method for conveying intent (C1-C6). \textit{``Particle textures and colors work well with semantic descriptions, but shapes and trajectories are better suited to graphical representation''} (C6). 
We observed that C2 lacked a clear semantic concept when designing the ``impact'' effect but had a well-defined approach to motion design. Conversely, C3 took the opposite approach when designing ``lightning''. This suggests that semantic input and graphical input should support flexible prioritization between the two.
Additionally, creators emphasized that graphical input should be intuitive and simple. \textit{``I want to quickly sketch the shapes and paths in my mind through straightforward operations''} (C5).
\label{F2}

\Ffinding{\textbf{Intermediate results are beneficial for expressing preferences and inspiring ideas.}}
In Task 3, creators commonly began by inputting semantic queries to search for particle effects, then iteratively refined their inputs to explore more satisfactory results. For example, C3 experimented with replacing terms like ``element,'' ``energy,'' and ``lightning'' in their search query before finding a desirable outcome. However, we identified an alternative and more effective exploration method: leveraging intermediate results to express user preferences.
Creators expressed a strong preference for building upon moderately satisfactory results rather than repeatedly modifying input queries (C1, C3). For particle effects, demonstrating preferences through examples proved more accurate and less cognitively demanding than explicit semantic descriptions. \textit{``It's easy to point to an effect I like, but much harder to describe what I want''} (C4).
Additionally, intermediate results helped stimulate design inspiration and clarify creative intent. During exploration, creators often encountered unexpected yet appealing effects that diverged from their initial intent (C1, C2, C3, C4). These discoveries could spark new design directions—\textit{``I originally wanted straight trajectories, but after seeing spiral ones, I preferred those instead''} (C1). Visual feedback from intermediate results also assisted in rapidly converging vague ideas: \textit{``I wasn’t sure how the lightning’s path should look, but one of the effects I found worked well—I’d like to explore more variations of that trajectory''} (C3).
\label{F3}

\Ffinding{\textbf{Creating innovative works by transferring particle effects and combining them.}}
After the task, we reviewed the results with the creators. We found that some particle effects could be effectively transferred to different scenarios to meet their needs. For example, C4 repurposed an explosive ``flash'' effect to convey ``broken,'' while C5 noted that a horizontal ``ring'' effect originally designed for skill ranges could express a ``portal'' effect when rotated vertically. C6 pointed out that a ``spark'' particle effect could represent different phenomena—such as embers, electric sparks, or water splashes—simply by altering its color.  
We observed that, based on semantic or kinematic similarities, many particle effects could be adapted to other contexts through modifications like color changes, translation, rotation, or scaling, expanding creative possibilities. Another approach to enhancing creativity was the recombination of particle effects. Our dataset consisted of composite effect works assembled from individual particle effects. \rv{Novel effect artworks beyond the original themes of the particle effect dataset have the potential to be created through transferring and recombining these effects.}
However, creators expressed concerns about whether the combinations of explored particle effects would harmonize well (C3, C6). This insight highlighted the need to ensure that features like color, shape, and size remain coordinated during the exploration process, facilitating the creation of visually cohesive compositions.
\label{F4}

\subsection{Qualitative Analysis}
From semi-structured interviews, we identified that kinematic behavior of particle effects is both crucial and challenging to describe by semantics. To better understand these kinematic features and develop more efficient representation methods, we conducted a qualitative analysis of the 839 particle effects collected in Sec.~\ref{sec:asset}. Our analysis focused on the following dimensions of particle effects:

\begin{itemize}[leftmargin=0pt, itemindent=*]
    \item \textbf{Duration}: The time from when a particle effect appears until it completely disappears.
    \item \textbf{Particle Size}: The average size of particles during their lifetime.
    \item \textbf{Emission Shape}: The approximate visual shape of the region where particles are emitted. 
    \item \textbf{Emission Trail}: The emergent shape of particles after emission.
    \item \textbf{Symmetry}: The visual symmetry exhibited by the particle effect.
    \item \textbf{Kinematic Consistency in Effect Composition}: The correlation between individual particle effects within a composite work, based on the above features.
\end{itemize}

\begin{figure*}[t]
 \centering 
 \includegraphics[width=\linewidth]{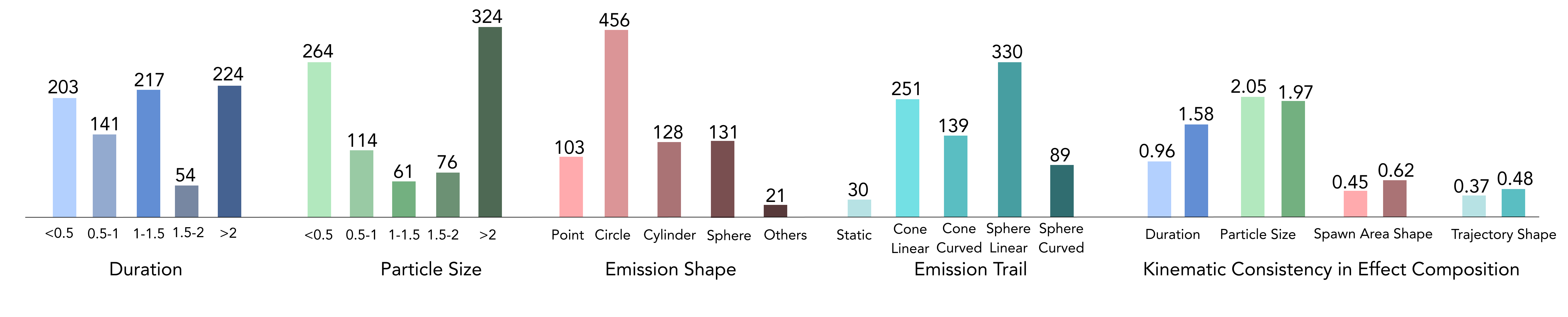}
 \caption{The distribution of particle effects across key dimensions.}
 \label{fig:distribution}
 \Description{Distribution Analysis}
\end{figure*}

These dimensions were selected due to their strong visual impact, as emphasized by creators and previous studies~\cite{raappana2018great, vaaraniemi2016example}. We analyze these dimensions to design a kinematic representation and distance metric that more effectively aligns with human visual perception. The results of our analysis are presented in Fig.~\ref{fig:distribution}.

\begin{figure}[tb]
 \centering 
 \includegraphics[width=\columnwidth]{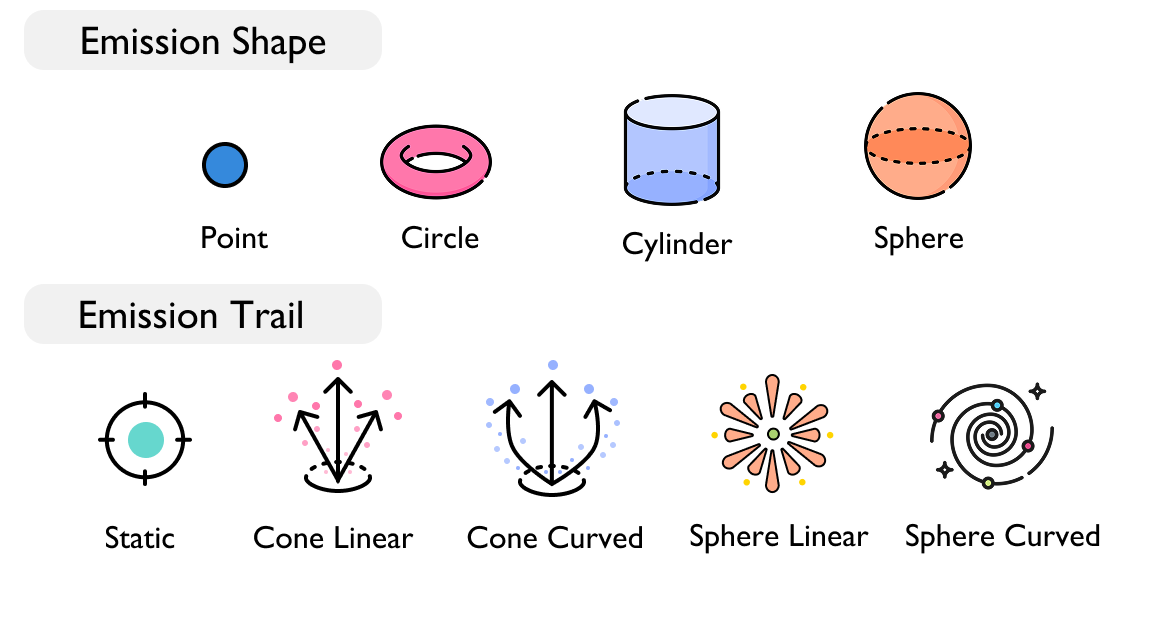}
 \caption{Common cases of particle effects in terms of emission shape and emission trail.}
 \label{fig:shape}
 \Description{shape}
\end{figure}

\textbf{Duration}.
The duration of particle effects spans a wide range. Effects with short durations typically exhibit instantaneous appearances, while those with long durations often demonstrate stable, sustained effects. We need to account for duration when representing particle effects.

\textbf{Particle Size}.
Particle sizes in effects vary significantly. Effects with small particles appear as distributed points, while those with larger particles display clearer textures. We need to consider particle size when representing effects.

\textbf{Emission Shape}.
The most common emission shapes for particles in effects are: Point (we consider areas with radius <0.1 as points), Circle, Cylinder, and Sphere (Fig. ~\ref{fig:shape}). Our representation of particle effects primarily considers these primitive shapes for emission regions.

\textbf{Emission Trail}.
We classify emission trails based on emission direction (conical emission along an axis or spherical emission) and trajectory shape (linear or curved) (Fig.~\ref{fig:shape}). Our representation of particle effect emission trails focuses on these common cases.

\textbf{Symmetry}.
While particle behavior during runtime exhibits randomness, the underlying distribution follows predictable patterns. We primarily consider the axial symmetry of these visual distributions. Among 839 samples: 817 (97.4\%) demonstrated axial symmetry along at least one axis, 18 (2.1\%) had non-symmetric emission regions, and 4 (0.48\%) were static images with asymmetric patterns. Therefore, our representation mainly considers axially symmetric cases.

\textbf{Kinematic Consistency in Effect Composition}.
We compared kinematic feature differences between multiple effects within the same artwork versus across different artworks for the first four dimensions. Distance metrics were calculated as:
1) Duration/Particle Size: Absolute value difference; 2) Emission Shape: 0 for identical values, 1 otherwise; 3) Emission Trail: 1 if both direction and shape differ, 0.5 if one matches, 0 for identical.
Results show significantly smaller (significance level $p < 0.001$) feature differences (duration, emission shape, emission trail) within artworks than between artworks. This suggests that effects with similar kinematic behavior characteristics can be more harmoniously composed. Our proposed method incorporates this insight.

\subsection{Design Goals}
Informed by our formative study, we outlined four goals for the design and development of KinemaFX, aiming to support non-expert users in flexibly expressing their intent and efficiently exploring and customizing particle effects.

\Ggoal{\textbf{Efficient representation and measurement of kinematic behavior (F\ref{F1}).}}
To address the limitations of semantic descriptions and LLM-based perception of kinematic behavior, our method should provide a more efficient and accurate way to represent the kinematic behavior of particle effects. Additionally, it needs to support distance metrics between different kinematic behaviors to facilitate similarity comparisons among particle effects. \label{G1}

\Ggoal{\textbf{Flexible intent expression combining semantic and graphical inputs (F\ref{F2}).}}
Our approach should accommodate both semantic and graphical inputs to help users better articulate their intent while reducing mental load. The balance between these input modalities should be adjustable, allowing users to flexibly prioritize either semantic or kinematic aspects based on their needs. \label{G2}

\Ggoal{\textbf{Supporting preference-guided exploration through intermediate results (F\ref{F3}).}}
The system should leverage user feedback on intermediate results to refine preferences and dynamically guide the exploration process in an iterative manner, adapting to evolving user intent. \label{G3}

\Ggoal{\textbf{Coherent transfer and composition of particle effects (F\ref{F4}).}}
To facilitate the creation of novel effects, the method should support seamless adaptation of particle effects to different scenarios through simple transformations, as well as efficient composition. These operations should ensure visual coherence to produce satisfying effect artworks. \label{G4}

\section{KinemaFX}
Based on the design goals, we proposed KinemaFX, a kinematic-driven interactive system to assist non-expert users in constructing customized particle effect artworks.

\subsection{Method Overview}

\rv{Informed by the formative study and to support efficiently representing complex kinematic behavior in human-AI collaboration, we propose a conceptual model of particle effects. Then we developed KinemaFX based on the conceptual model.}

\rv{
\subsubsection{Conceptual Model}

Informed by our formative study, we propose \textit{a conceptual model} to describe particle effects. Particle effects typically contain both semantic aspects that are easily expressed through language, and kinematic characteristics that are more naturally described through graphical or mathematical representations. For example, for a ``fire burst'' effect, terms like ``flame'', ``magical'', or ``realistic'' can effectively convey the style, color, and texture, while graphical representations (e.g., lines or primitive shapes) can be more intuitive and accurate for describing kinematic properties such as the burst’s spread range, angle, trajectory, and duration.

Therefore, our conceptual model explicitly separates these two components to capture both the semantic features and kinematic behavior of particle effects. \textit{The semantic component} encodes the intended meaning and stylistic qualities, while \textit{the kinematic component} captures the spatiotemporal structure and particle motion. Given that particle effects are homogeneous, multi-object, three-dimensional, and dynamic in nature, their kinematic behaviors are particularly challenging to describe in a user-friendly way.
To address this, our core idea is to represent the motion of multiple particles as the transformation of a composite shape. Building on this idea and guided by an analysis of particle effect assets, we decompose kinematic behavior into three key components: duration, emission shape, and emission trail. These elements respectively describe the temporal length, initial spatial structure, and dynamic transformation of the effect. This formulation strikes a balance between expressiveness and usability, capturing complex kinematic features while remaining intuitive for interaction and precise for design control.
}

\rv{
\subsubsection{Kinematic-driven System Design}

Based on the conceptual model we proposed, we developed KinemaFX in a kinematic-driven approach.
Kinematic-driven means that complex kinematic information is fully and precisely captured and communicated throughout the human-AI collaborative design process. It is characterized by two key features: (1) a structured representation that intuitively and comprehensively encodes kinematic information, ensuring fidelity and completeness during human-AI interaction; and (2) support for efficient computation and comparison, allowing for iterative refinement of kinematic features to enable fine-grained, precise control.
}

KinemaFX enables users to flexibly express their intent and efficiently explore the design space of particle effects through implicitly expressed preferences. It consists of a workflow for particle effect exploration and customization, as well as a method to enhance particle effect search within the workflow. We present an overview of KinemaFX in Fig.~\ref{fig:teaser}.

\textbf{Implicit user preference-guided exploration workflow.}
We allow users to flexibly express their intent through a combination of semantic and graphical inputs \textbf{(G\ref{G2})}. Then, we support users in exploring particle effects in an iterative process: in each round, users select particle effects of interest, and KinemaFX alternately conducts searches either within the neighborhood of user-selected intermediate results or along directions relevant to user preferences. This iterative exploration is guided by implicit preferences expressed through the user’s intermediate selections \textbf{(G\ref{G3})}. Finally, we enable users to refine and combine the particle effects selected during exploration to create an effect artwork \textbf{(G\ref{G4})}. The details of the workflow are elaborated in Sec.~\ref {sec:workflow}.

\textbf{Enhancing particle effect search and alignment through structured representation and distance metric.}
We propose a two-stage method to address the limitations of representing and comparing the kinematic behavior of particle effects semantically or via LLMs, thereby supporting particle effect search and alignment during the iterative exploration phase of the workflow. 
In stage 1, we introduce a structured representation for combining semantic and kinematic behavior \textbf{(G\ref{G1})}. Specifically, we abstract the kinematic behavior of particle effects as dynamic changes of primitive shapes and represent them as mathematical expressions in a spherical coordinate system. Then in stage 2, we leverage this representation to enable efficient difference measurement between instances and promote coordination among particle effects through kinematic behavior alignment \textbf{(G\ref{G4})}. The two-stage method is illustrated in Fig.~\ref{fig:kinematic} and we present the technical details in Sec.~\ref{sec:kinematic}.


\subsection{Workflow}\label{sec:workflow}
The workflow of KinemaFX consists of three stages: initial intent input, iterative exploration guided by implicit user preference, and effect composition. Within this workflow, the iterative exploration alternates between two exploration modes: local exploration and directional exploration.

\subsubsection{Initial Intent Input}

\begin{figure}[tb]
 \centering
 \includegraphics[width=\columnwidth]{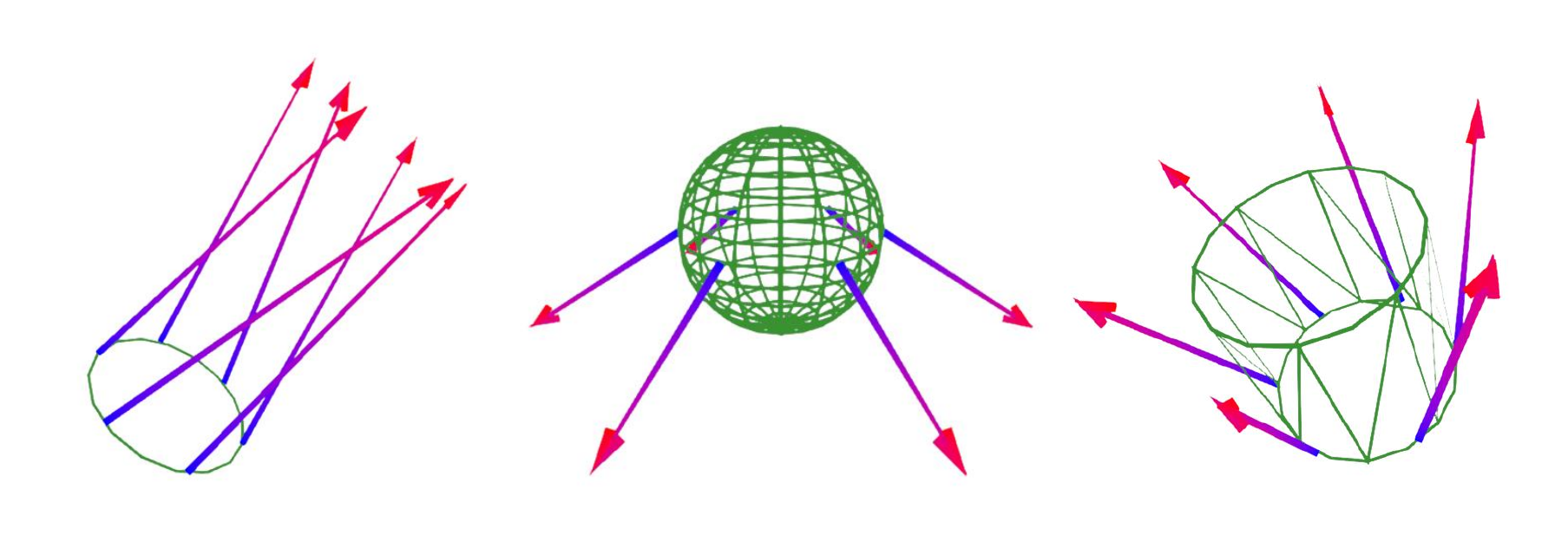}
 \caption{Examples of graphical input. Users can intuitively express kinematic-related intentions through graphical input. The emission shape and emission trail can be flexibly manipulated.}
 \label{fig:input}
 \Description{Graphical Input}
\end{figure}

We have found that the kinematic behavior of particle effects is crucial yet difficult to effectively and intuitively describe using semantics. Users expressed the need for both semantic and graphical input, as well as the flexibility to adjust the emphasis between the two when articulating their design intent for particle effects. Accordingly, KinemaFX supports a flexible combination of semantic and graphical input for intention expression.

For semantic input, users describe the particle effects using natural language in a text box. \rv{User input in natural language will be processed by the LLM into more standardized and concise semantic descriptions of particle effects, facilitating subsequent conversion into embeddings.} For graphical input, our design aims to enable natural and quick representation of the approximate kinematic behavior of particle effects without requiring specialized skills. Our system design is based on the graphical elements used by creators during interviews to express their intent, as well as the analysis of particle effect assets. We focus on three key features to describe kinematic behavior: duration, emission shape, and emission trail. The values of these features are constrained within reasonable ranges, and users can control them through common interaction methods:

\begin{itemize}[leftmargin=0pt, itemindent=*]
    \item \textbf{Duration}: The time from when a particle effect appears until it completely disappears. Controlled via a slider.
    \item \textbf{Emission Shape}: We consider three basic shapes (circle, sphere, and cylinder) and control their transformations through translation, rotation, and scaling of 3D shapes. We do not consider the point shape here, as it essentially represents the aforementioned shapes with minimal radius values.
    \item \textbf{Emission Trail}: Represented by axisymmetric vectors, controlled via mouse-drawn lines. Given the difficulty of directly positioning points in 3D space with a mouse, we allow users to drag to control the projection of the vector endpoint onto the plane formed by the vector and the axis of symmetry, while a rotation axis controls the spiral of the trajectory.
\end{itemize}

The examples of graphical input in KinemaFX are shown in Fig.~\ref{fig:input}. To support users in adjusting the emphasis between semantic and graphical input when expressing intent, we provide a slider to control the weighting of their combination. Through this input method, users articulate their intent as a weighted integration of semantic information and structured graphics, which serves as input for the subsequent iterative exploration phase.

\begin{figure*}[t]
 \centering
 \includegraphics[width=\linewidth]{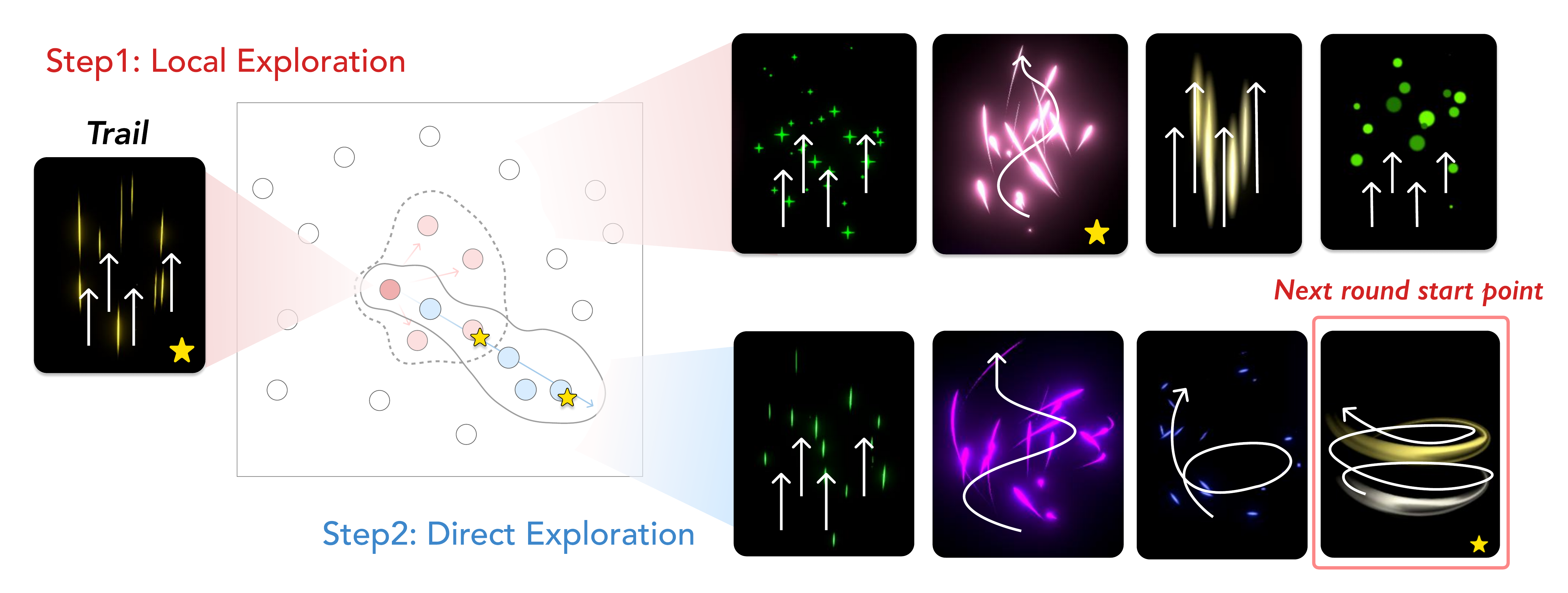}
 \caption{This case demonstrates the interplay between local and directional exploration: In Step 1, the user begins with a particle effect featuring a vertically rising trail. Through local exploration, they discover similar effects and select one with a spiraling upward trail. Then, in Step 2, directional exploration is performed based on the initial and newly selected effects. This yields a set of particle effects that preserve shared features including the trail semantics and upward movement, while introducing variations in the degree of spiraling.}
 \label{fig:exploration}
 \Description{}
\end{figure*}

\subsubsection{Local Exploration}
The purpose of local exploration is to find the particle effects that best match the user's current intent. Initially, the user's input intent serves as the starting point for local exploration in the iterative exploration phase. Local exploration searches the particle effect database based on the input semantic and kinematic information, as well as the user-controlled weighting between them, returning the top K most similar results (K is set to 4 by default in our system). The user then selects the most satisfactory result from these K options to proceed with further exploration. \rv{The similarity of particle effects is calculated as a weighted sum of the semantic similarity and the kinematic similarity, based on user-adjusted weights. The semantic similarity is computed using the cosine similarity of embeddings, while the calculation of the kinematic similarity will be detailed in Sec.~\ref{sec:search}.}
Non-expert users often struggle to articulate their intent precisely through semantic input or manipulate complex graphical inputs and repeatedly adjusting intent expression to explore different results also increases mental load. To address this, KinemaFX uses the particle effects selected by the user in each iteration as intermediate results to represent their evolving intent. The specific representation of particle effects used for searching similar results remains transparent to users, thereby eliminating the need for them to bear the cognitive load associated with learning and explicitly expressing their preferences. As users iteratively select preferred effects and continue exploring similar results, this progressive process is guided by their implicitly expressed preferences through intermediate selections. This approach aligns with the user's intent in real time while reducing the mental load of explicit preference expression. We illustrated the process of local exploration in Step 2 of Fig.~\ref{fig:exploration} and the implementation details of local exploration are elaborated in Sec.~\ref{sec:kinematic}.

\subsubsection{Directional Exploration}
The purpose of directional exploration is to expand the search along the direction of the user's implicitly expressed preferences. Relying solely on local exploration may confine users to a small region of the design space near their initial intent, leading to local optima. Prior works~\cite{suh2024luminate, almeda2024prompting} prevent this by conducting structured searches across the design space, but such methods fail to account for user intent, making them ineffective for high-dimensional or large design spaces.
KinemaFX addresses this through directional exploration. The input of local exploration is the particle effect the user found interesting in the previous step, and the newly selected effect is their current preference. The direction of change between these two effects is interpreted as the user's preferred exploration direction. We then search for new effects at varying distances along this direction to present potential results for the next step.
Directional search preserves shared features (semantic and kinematic) between the two reference effects while exploring diverse values for differing features. The LLM infers multiple particle effect descriptions that share common semantic features with the two input effects while extending variations at the points of difference. Also, we explore additional kinematic behaviors along the target direction by extrapolating different extrapolation coefficients between the two effects' kinematic representations. Directional exploration allows the exploration process to be guided by implicit user preferences, enabling more efficient coverage of the design space and preventing premature convergence to suboptimal solutions. Implementation details are provided in Sec.~\ref{sec:kinematic}. Fig.~\ref{fig:exploration} illustrates an example of how local exploration and directional exploration alternate during iterative exploration to support efficient particle effect search.

\subsubsection{Effect Composition}
The goal of effect composition is to enable users to assemble collected particle effects from the iterative exploration phase into a cohesive effect artwork. To ensure visual harmony, adjustments to position, orientation, scale, duration, playback speed, and start delay are often necessary.
To minimize manual adjustments, KinemaFX incorporates kinematic behavior alignment during searching, reducing post-processing effort. Users can thus efficiently combine satisfactory particle effects into novel effect artworks. KinemaFX supports exporting completed effect artworks.

\subsection{Particle Effect Search and Alignment}\label{sec:kinematic}

\begin{figure*}[t]
 \centering
 \includegraphics[width=\linewidth]{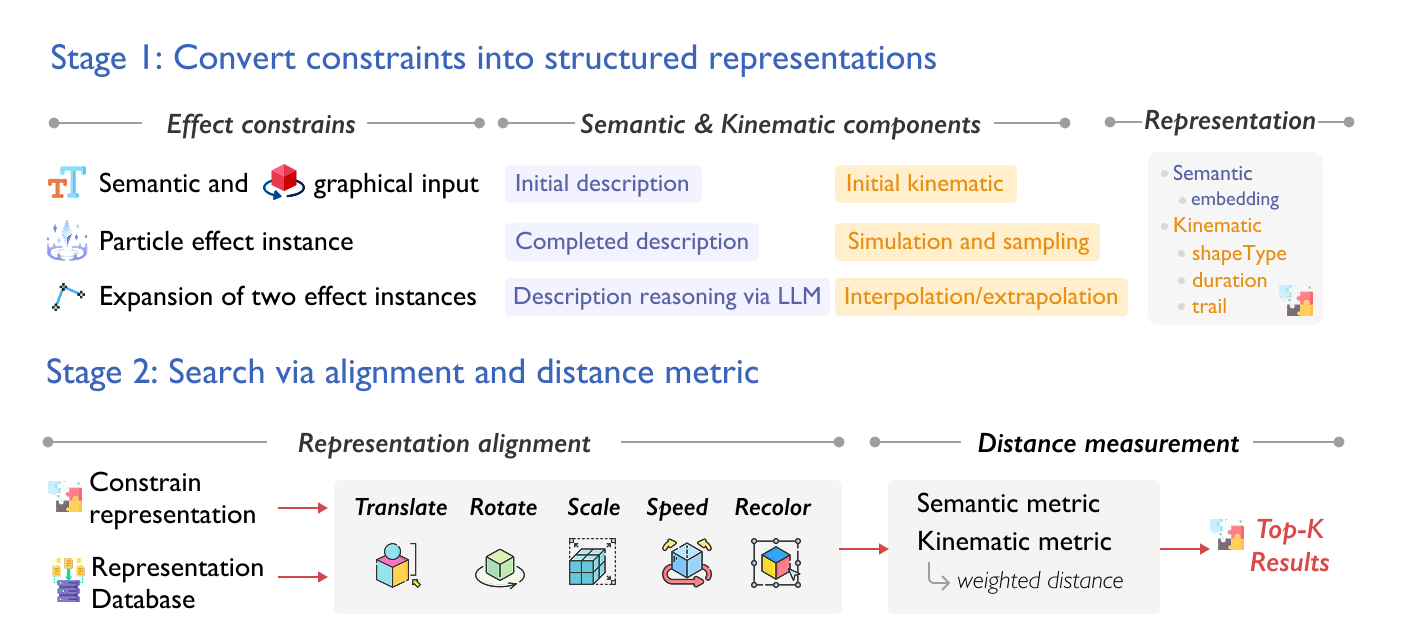}
 \caption{We propose a two-stage kinematic-driven method to enhance particle effect search and alignment: convert constraints into structured representations and search via alignment and distance metric.}
 \label{fig:kinematic}
 \Description{}
\end{figure*}

To enhance particle effect search and alignment, we propose a two-stage kinematic-driven method \rv{based on our conceptual model}: In stage 1, we propose a structured representation method for particle effects and support transforming different search constraints into structured representations. In stage 2, based on the structured representations from stage 1, we support efficient particle effect search through alignment and distance metrics. Fig.~\ref{fig:kinematic} illustrates our two-stage method.

\subsubsection{Structured Representation}

A particle effect is represented as \( R = (S, K) \), where \( S \) denotes the semantic component and \( K \) represents the kinematic behavior. 
\rv{For the semantic component, we use the semantic description of the effect processed by the sentence-transformers model to obtain the embedding.}

For the kinematic behavior, based on the findings of our formative study, we model the aggregated motion of particles over time as the translation, rotation, and scaling of an overall geometric shape. This modeling leverages the axisymmetric property commonly observed in particle effects.
We define the kinematic behavior as:
\begin{equation}
K = (\text{shape}, \text{trail}, \text{duration}),
\end{equation}
where \( \text{shape} \) describes the emission shape of the particle effect, \( \text{trail} \) represents the temporal transformation of the shape over time, and \( \text{duration} \) denotes the total duration of the particle effect.
The shape is parameterized as:
\begin{equation}
\text{shape} = (s, r, h),
\end{equation}
where \( s \in \{\text{circle}, \text{cylinder}, \text{sphere}\} \) indicates the type of shape, \( r \) are the outer radii, and \( h \) is the height (applicable to cylinders).
For the trail, we map the shape's trajectory to the spherical coordinate changes of the particles at the outer boundary points of the initial shape. This allows the kinematic behavior of the particle effect to be expressed mathematically in spherical coordinates. Based on the axisymmetric property of particle effects (typically along the vertical axis), we align the axis of symmetry of the particle effect with the polar axis of the spherical coordinate system. The spherical coordinates of a particle on the outer boundary of the shape are denoted as \( p = (r, \theta, \phi) \), where the translation corresponds to the change in \( r \cdot \cos(\theta) \), the rotation corresponds to changes in \( \phi \), and the scaling corresponds to the distance between the projection point on the equatorial plane and the pole, \( r \cdot \sin(\theta) \).

Based on this axisymmetric property, we can describe the shape evolution using the changes in the spherical coordinates of the initial shape’s outer boundary particles. The trajectory of the shape evolution is represented as:
\begin{equation}
\text{trail} = \{(\Delta r_i, \Delta \theta_i, \Delta \phi_i)\}_{i=1}^{N},
\end{equation}

where \( N \) is the number of divisions for the particle’s lifetime (default value of \( N = 8 \) in our system), and \( (\Delta r_i, \Delta \theta_i, \Delta \phi_i) \) represents the change in spherical coordinates of the boundary particles in the \( i \)-th time period of the particle’s lifecycle. For the duration, we define it as the time from the first particle generation to the last particle's dissipation.

\subsubsection{Convert Constraints into Structured Representations}\label{sec:convert}

We introduced our approach for this stage in Stage 1 of Fig.~\ref{fig:kinematic}.
Our particle effect search method represents search constraints as structured representations of particle effects, enabling direct matching with structured representations of effects stored in the database. To support this process, KinemaFX accommodates three types of constraints that can be transformed into structured representations:
1) User's semantic and graphical input – used for the search based on the user's input.
2) A single particle effect instance – used for local exploration and for converting existing particle effects from the database into structured representations.
3) Extrapolation based on two particle effect instances – used for directional exploration.

For user input transformation, \rv{we first process the user's natural language input into more standardized and concise semantic descriptions of particle effects using the GPT-4o-mini model, and then generate embeddings of the semantic descriptions using the sentence-transformers model all-MiniLM-L6-v2.} The graphical input, including shape, trail, and duration, directly corresponds to the kinematic representation format, allowing easy conversion.

\begin{figure}[tb]
 \centering
 \includegraphics[width=\columnwidth]{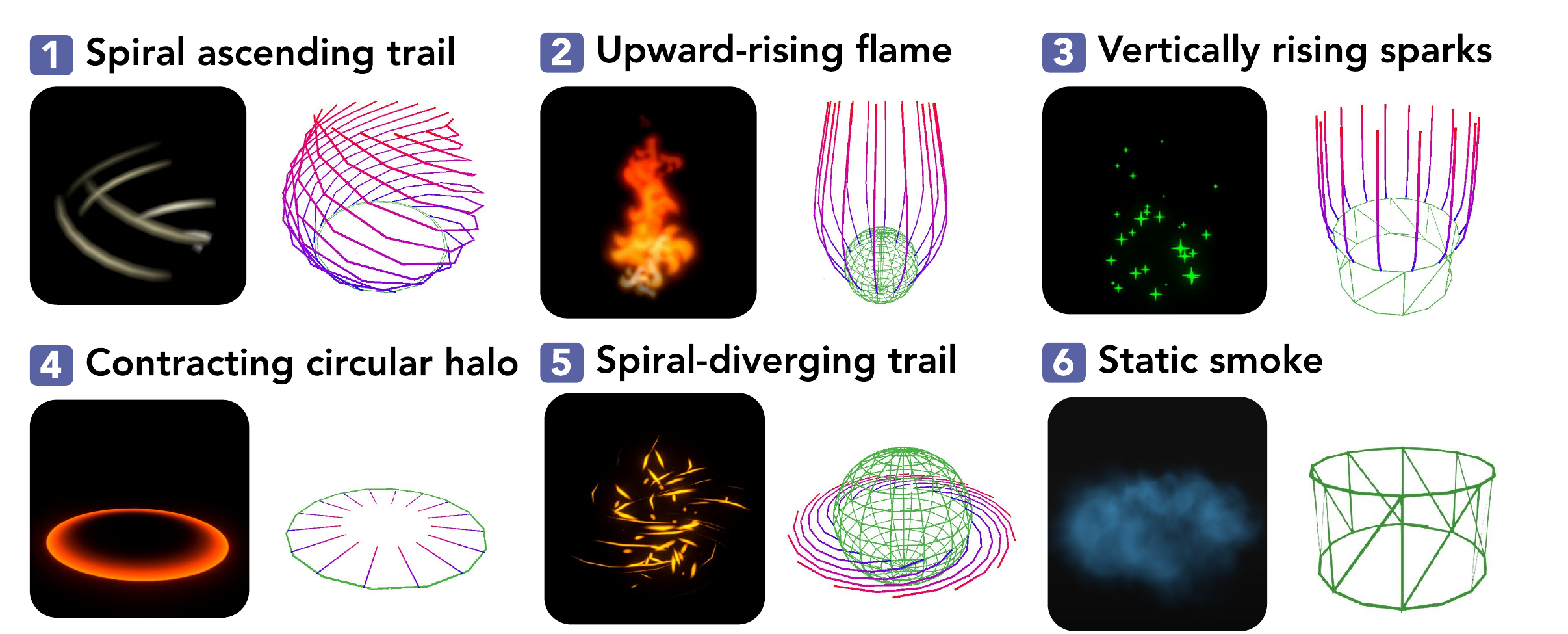}
 \caption{Example particle effects with different kinematic behaviors and their visualized kinematic representations. The kinematic representations of the particle effects preserve the essential kinematic information.}
 \label{fig:kinematicExmaples}
 \Description{}
\end{figure}

For transforming a particle effect instance, we construct the semantic representation by concatenating the textual description of the effect with its associated artwork description and encoding it into an embedding. The kinematic representation is obtained by simulating the particle effect, sampling all particles over their lifetimes, and extracting positional, size, and orientation data at multiple time steps. These sampled properties are then converted into spherical coordinates and averaged to derive the kinematic representation parameters, ensuring a visually consistent structured representation of the particle effect’s kinematic behavior. Fig.~\ref{fig:kinematicExmaples} shows several particle effects with different kinematic behaviors and their kinematic representation visualized.

For expansion based on two particle effect instances, the LLM analyzes the commonalities and differences between the two effects and generates multiple candidate descriptions of particle effects sharing similar properties while extending variations at the points of difference. \rv{The user's initial semantic input is incorporated into the semantic information of the particle effect as a constraint to ensure relevance to the user's original intent.} The kinematic representation is computed through extrapolation between the two effects. Specifically, the shape and the trail are extrapolated based on three spherical coordinate components: projection along the polar axis, rotation around the polar axis, and projection onto the equatorial plane relative to the pole. Duration is extrapolated directly. To ensure the extrapolation remains within a valid range, all extrapolations are performed logarithmically. By combining the semantic results from the LLM and the extrapolated kinematic representation, we obtain the structured representation for directional exploration.

\subsubsection{Search via Alignment and Distance Metric}\label{sec:search}

We introduced our approach for this stage in Stage 2 of Fig.~\ref{fig:kinematic}.
To retrieve proper particle effects from the database, we compute the similarity between the structured representation of a search constraint and those of existing effects, ranking the results accordingly. The search process incorporates transformations—translation, rotation, scaling and duration adjustment—to optimally align particle effects with the given constraint. Formally, we define the search objective as:

\begin{equation}
    R^* = \arg\min_{R_i \in \mathcal{D}, T \in \mathcal{T}} D(R_c, T(R_i)),
\end{equation}

where \(R_c\) represents the structured representation of the particle effect search constraint, and \(R_i\) represents the structured representation of a particle effect from the database \(\mathcal{D}\). The transformation \(T\) is applied to \(R_i\), and \(D(R_c, R_i)\) measures the distance between the two structured representations. This distance is composed of two components: the semantic distance \(D_s(S_c, S_i)\), which is computed via the cosine similarity of their embeddings, and the kinematic behavior distance \(D_k(K_c, K_i)\), which quantifies the variation in shape evolution over time.

To compute \(D_k(K_c, K_i)\), we first apply the transformations along the trail to the shape at each discrete time step \(t_j\) (where \(j = 1, \dots, N\)), yielding transformed shapes \(\text{shape}_c^{(j)}\) and \(\text{shape}_i^{(j)}\). The discrepancy at each time step is evaluated using the Hausdorff distance \(H(\cdot, \cdot)\), which captures shape dissimilarities, supplemented by a rotation penalty term:

\begin{equation}
    D_{\text{shape}}^{(j)} = H(\text{shape}_c^{(j)}, \text{shape}_i^{(j)}) + \lambda (1 - \cos(\Delta\phi_j)).
\end{equation}

Here, \(\text{shape}_c^{(j)}\) and \(\text{shape}_i^{(j)}\) are the shapes of the particle effects at time step \(t_j\) after applying the transformations. \(H(\cdot, \cdot)\) is the Hausdorff distance, which measures the dissimilarity between two shapes. The parameter \(\lambda\) controls the weight of the rotation penalty, and \(\Delta\phi_j\) represents the rotation difference between the shapes at time step \(t_j\).

The overall trail distance is computed by summing the shape distances across all time steps:

\begin{equation}
    D_{\text{trail}}(\text{trail}_c, \text{trail}_i) = \sum_{j=1}^{N} D_{\text{shape}}^{(j)}.
\end{equation}

Additionally, we introduce a non-linear duration adjustment factor to measure differences in the visual persistence of particle effects. This factor minimizes the impact when the durations are similar but amplifies the differences when the durations are significantly different. The duration adjustment factor is modeled as:

\begin{equation}
    f(\text{duration}_c, \text{duration}_i) = 1 + \alpha \left( \max \left( \frac{\text{duration}_c}{\text{duration}_i}, \frac{\text{duration}_i}{\text{duration}_c} \right) - 1 \right)^2,
\end{equation}

where \(\text{duration}_c\) and \(\text{duration}_i\) are the visual durations of the search constraint and the particle effect \(P_i\), respectively, and \(\alpha\) is a tunable parameter that controls the weight of the duration difference.

The final kinematic behavior distance is given by:

\begin{equation}
    D_k(K_c, K_i) = D_{\text{trail}}(\text{trail}_c, \text{trail}_i) \cdot f(\text{duration}_c, \text{duration}_i).
\end{equation}

By combining the semantic and kinematic behavior distances with appropriate weighting, the search method effectively retrieves and ranks particle effects that are visually and semantically similar to the user's query. The incorporation of transformation alignment further enhances the searching accuracy, ensuring the selected effects maintain high perceptual relevance.

\begin{figure*}[t]
 \centering
 \includegraphics[width=\linewidth]{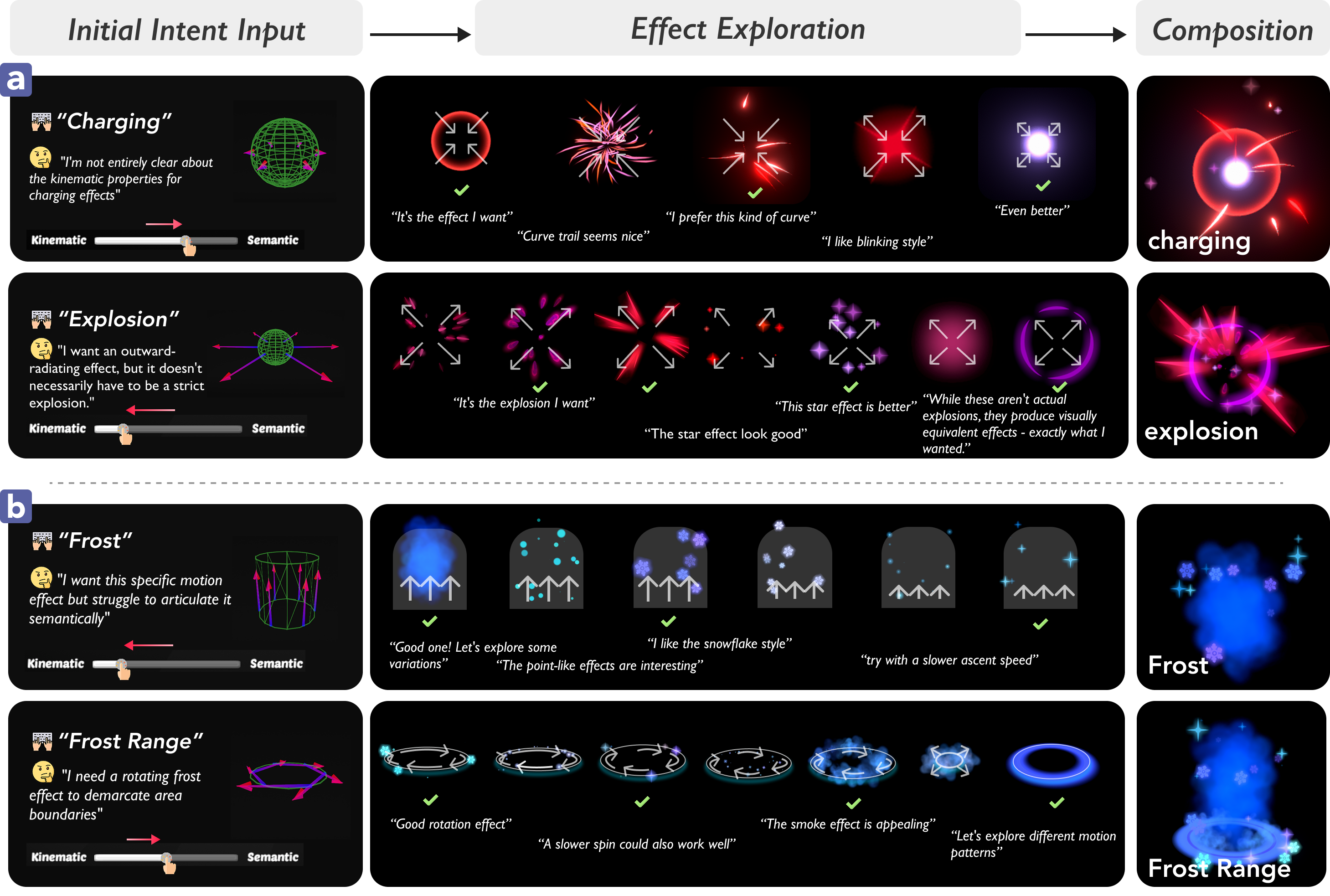}
 \caption{We demonstrate the usage of KinemaFX through two usage scenarios. (a) ``Charge-explosion'' effect: The user divided this effect temporally into two parts—``charging'' followed by ``explosion.''  
(b)``Frost range'' effect: The user divided this effect spatially into two parts—an aerial frosty atmosphere and a frost-covered ground area.  
In both scenarios, the user explored the different parts separately and then composed them into coherent effect artworks.}
 \label{fig:case}
 \Description{}
\end{figure*}

\subsection{Usage Scenarios}
We demonstrate how KinemaFX can be used to explore and create a particle effect artwork through two usage scenarios. The process is illustrated in Fig.~\ref{fig:case}.

\textbf{``Charge-explosion'' Effect.} The user divided this effect temporally into two parts: ``charging'' followed by ``explosion'', and explored them separately. For the charging stage, the user provided a brief semantic description and a vague spherical range as input. Since the kinematic behavior was unclear, the user steered the exploration to rely more on semantic guidance, and subsequently discovered a set of satisfying results. For the explosion stage, the user had a clearer sense of the desired kinematic behavior and relaxed the semantic constraints, resulting in effects that were not semantically labeled as explosions but shared visually similar dynamics. The user then selected several particle effects from the exploration process and adjusted their transformations and temporal features to compose a coherent ``charge-explosion'' effect artwork.

\textbf{``Frost range'' Effect.} The user divided this effect spatially into two parts: an aerial frosty atmosphere and a frost-covered ground area, and explored them separately. For the frosty atmosphere, the user could not think of a detailed semantic description beyond ``frost'', so the exploration focused on kinematic input and led to the discovery of stylistically diverse effects. For the frost-covered ground, the user balanced both semantic and kinematic inputs, resulting in multiple satisfying particle effects. The user then selected a subset of effects from the exploration and composed them by adjusting transformations and temporal features to create a complete ``frost range'' effect artwork.

\section{User Study}

We conducted a user study employing an ablation approach to evaluate two key modules in our system: implicit preference-guided exploration and the kinematic-driven search method.

\subsection{Method}

\textbf{Study Conditions.}
We adopted a within-group design in which each participant experienced four different experimental conditions. These conditions varied based on the presence or absence of the two key modules:

\begin{itemize}[leftmargin=0pt, itemindent=*]
    \item \textbf{Baseline}: Neither module is used. Participants express their intent through semantic input, and the system always recommends the particle effects that are semantically closest to the input.
    \item \textbf{Preference-Guided Exploration}: The kinematic-driven method is not used. Participants express their intent semantically, and the system supports iterative exploration based on the participants’ intermediate selections, guided by semantic information.
    \item \textbf{Kinematic-Driven Search}: The implicit preference-guided module is not used. The system combines semantic and kinematic information to search for the particle effect most aligned with the user's input, but does not support implicit preference-guided exploration.
    \item \textbf{KinemaFX}: Both modules are enabled, providing the full functionality of our system.
\end{itemize}

\rv{
\textbf{Datasets.}
The dataset used in our user study consists of particle effect assets collected in the formative study, including 839 particle effects used to compose 147 particle effect artworks.
}

\textbf{Participants.}
We recruited 16 participants (age: 19–29, 8 male and 8 female). None of the participants had a professional background in particle effects design or development. Six participants had experience with game engines, and all participants had a basic understanding of particle effects.

\begin{figure}[tb]
 \centering 
 \includegraphics[width=\columnwidth]{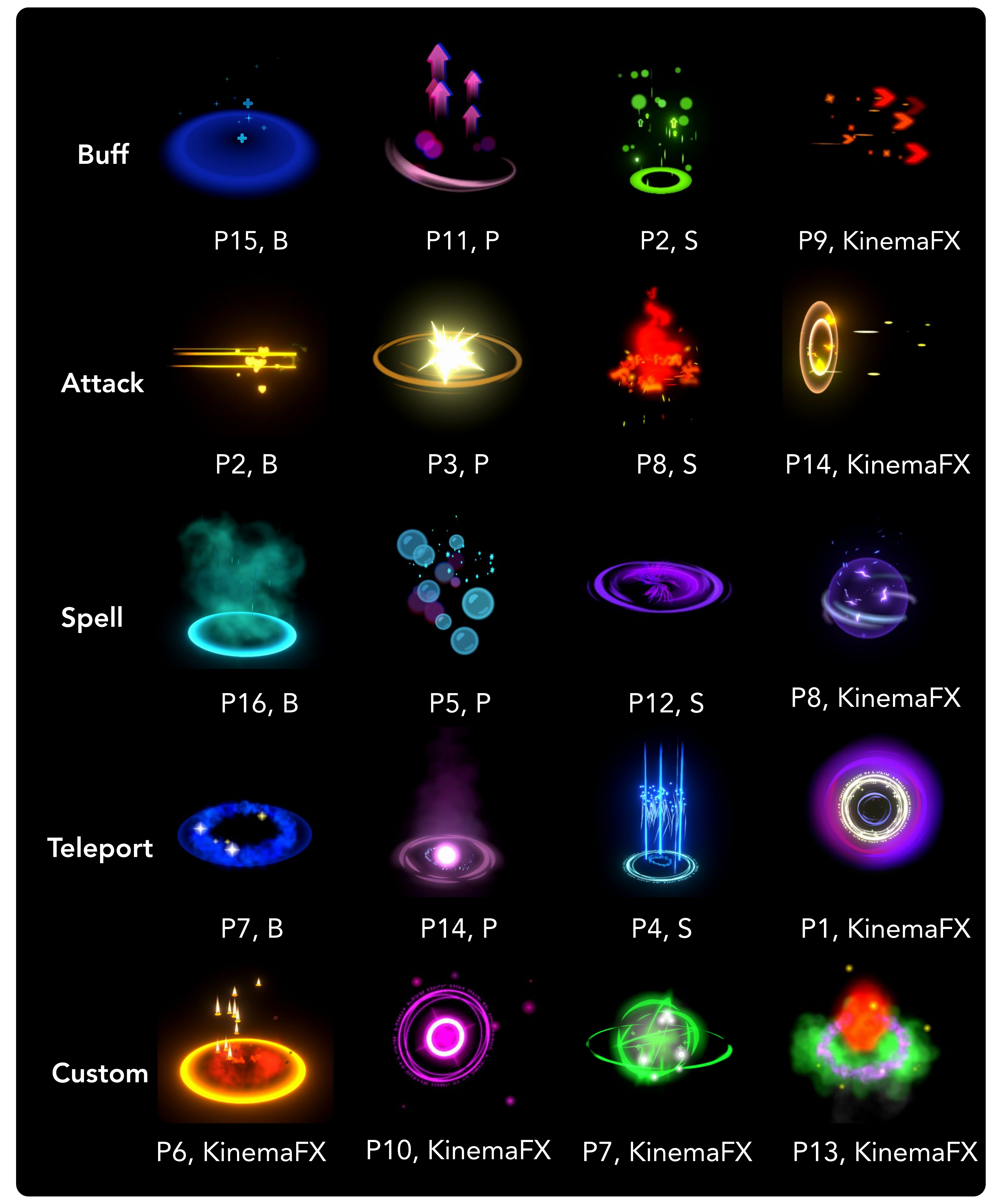}
 \caption{Representative effect artworks created by users under the four study conditions. B denotes Baseline, P denotes Preference-Guided Exploration, and S denotes Kinematic-Driven Search. Please see the supplementary video for dynamic effects and the complete collection of particle effects from our user study.}
 \label{fig:gallery}
 \Description{Gallery}
\end{figure}

\textbf{Procedure.}
Each participant was asked to create four particle effects under the four experimental conditions, with each effect corresponding to a distinct theme: Buff, Spell, Attack, and Teleport. \rv{Our primary focus was on comparing the different experimental conditions; the use of different themes aimed to avoid biases caused by a single theme or by participants creating the same theme under multiple conditions.} The order of conditions and the mapping between conditions and themes were counterbalanced across participants.
At the beginning of the study, researchers introduced the purpose and procedure, collected demographic information, and provided a tutorial on particle effects and system usage. Participants then completed the creation tasks for each condition, and were encouraged to think aloud as they worked. \rv{Since we mainly focus on the experience and outcome of the creation process, we did not impose any time constraints during the experiment.} After completing all four themed tasks, participants filled out a questionnaire and took part in a semi-structured interview. Finally, participants used the full version of KinemaFX to create a custom effect artwork. The entire session lasted approximately 65 minutes. Fig.~\ref{fig:gallery} showcases a gallery of particle effect artworks created by users.

\textbf{Measures.}
Based on our system design goals and prior research on LLM-assisted exploration tools~\cite{suh2024luminate, wang2024virtuwander} as well as NASA-TLX~\cite{hart1988development}, we designed a questionnaire covering 15 items across seven dimensions:
1) Intent Expression (Q1-Q2);
2) Convergent Exploration (Q3);
3) Divergent Exploration (Q4);
4) Kinematic Harmony (Q5-Q7);
5) Effect Artworks (Q8-Q10);
6) Effort (Q11-Q13);
7) Experience (Q14-Q15).
The questionnaire items are listed in Table.~\ref{tab:questionnaire}. Participants rated each item using a 7-point Likert scale (1: Strongly Disagree, 7: Strongly Agree). Additionally, we conducted semi-structured interviews to gather qualitative feedback on the creation experience under each condition. This helped us better understand how the two modules influenced participants’ approaches and outcomes in particle effect exploration and customization.

\begin{table}[ht]
\centering
\resizebox{\columnwidth}{!}{
\begin{tabular}{cl}
\toprule
\textbf{ID} & \textbf{Question} \\
\midrule
Q1  & I was able to clearly express my intent. \\
Q2  & I liked the input method for expressing intent. \\
Q3  & I was able to effectively influence the iteration direction. \\
Q4  & I discovered surprising and satisfying results. \\
Q5  & I was satisfied with the shape and motion of the search results. \\
Q6  & I didn’t need to manually adjust much when combining effects. \\
Q7  & The spatiotemporal relationship of effects was harmonious. \\
Q8  & I was satisfied with the effect artwork I created. \\
Q9  & The effect artwork contained the core features I intended. \\
Q10 & The final effect was diverse beyond the core features. \\
Q11 & I did not feel a heavy cognitive load. \\
Q12 & I did not feel a heavy physical load. \\
Q13 & The system was easy to learn. \\
Q14 & The system was easy to use. \\
Q15 & I would like to use this tool in the future. \\
\bottomrule
\end{tabular}
}
\caption{Questionnaire items (Q1–Q15).}
\label{tab:questionnaire}
\end{table}

\begin{figure*}[t]
  \centering
  \includegraphics[width=\linewidth]{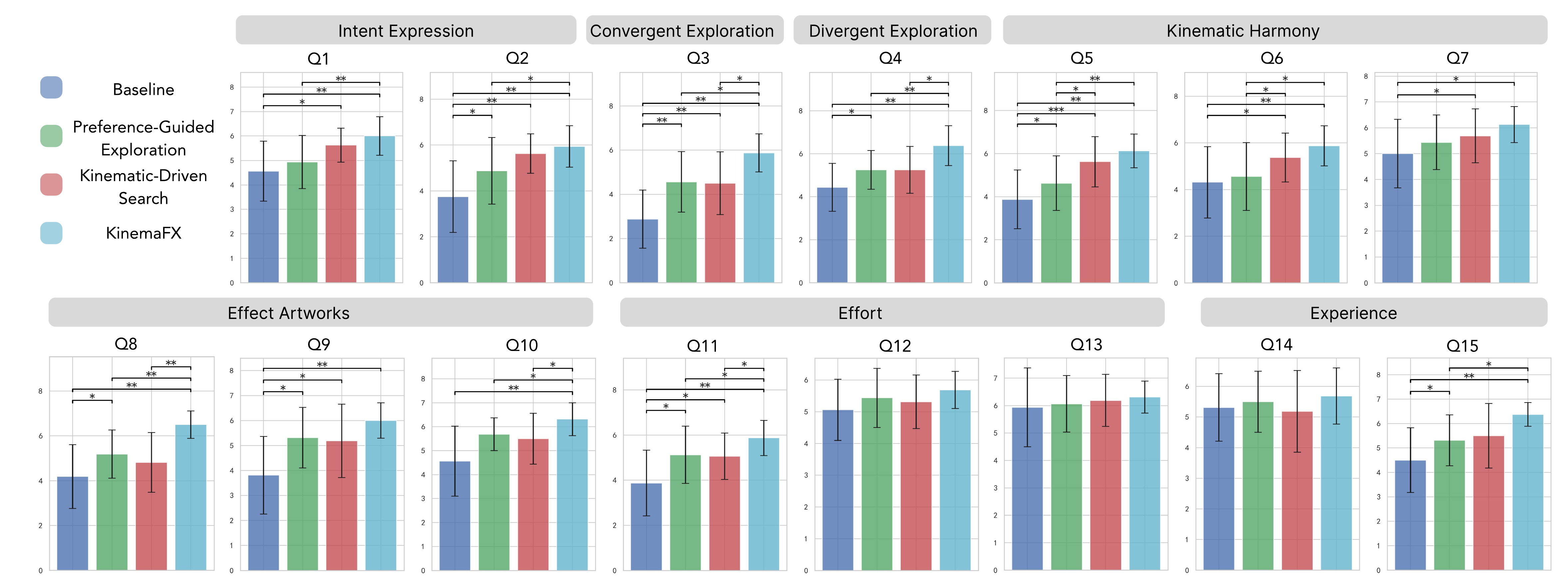}
  \caption{%
     Quantitative results of the user study. Significance levels are denoted as follows: (*) for $p < 0.05$, (**) for $p < 0.01$, and (***) for $p < 0.001$.
  }
  \vspace{-0.1in}
  \label{fig:results}
  \Description{Results}
\end{figure*}

\subsection{Results and Key Findings}

The primary analysis method was a combination of repeated measures statistical tests, including ANOVA~\cite{st1989analysis} or Friedman tests~\cite{Friedman01121937}, to compare the different conditions across various metrics. Pairwise comparisons between conditions were conducted to identify significant differences in performance. Bonferroni correction~\cite{weisstein2004bonferroni} was applied to account for multiple comparisons.
Significance levels are denoted as follows: (*) for $p < 0.05$, (**) for $p < 0.01$, and (***) for $p < 0.001$. The results are illustrated in Fig.\ref{fig:results}. Based on the results and qualitative feedback, we summarize the following key findings.

\textbf{The combination of semantic and graphical inputs allows for a more comprehensive and efficient expression of user intent.} 
We found that introducing graphical input benefits users in expressing their intent, which aligns with feedback from our formative study. 12 out of 16 participants mentioned that \textit{``describing motion trajectories using semantics is challenging,''} citing reasons such as \textit{``requiring a lot of text to describe details''} (P2), \textit{``inaccurate semantics''} (P5, P13), and \textit{``difficulty in finding the right words''} (P1, P3, P10), among others. P6 further noted that \textit{``adjusting visible graphics helps better determine one’s intent.''} This suggests that graphical elements can serve not only as input but also as visual feedback to help users clarify their intent.

\textbf{Expressing preferences through intermediate results better supports convergent and divergent exploration.}
We found that implicit preference guidance benefits exploration, both with and without the kinematic-driven search condition. First, selecting satisfactory instances to express preference better supports convergent exploration. P8 mentioned, \textit{``Maybe my initial input was not accurate, but once I found a satisfactory effect, I could explore more similar results in the next round.''} P12 noted, \textit{``I can just input an abstract concept and then clarify my intent further by selecting one of the search results.''} Meanwhile, directional exploration supported divergent thinking. Five participants stated that under such exploration conditions they \textit{``discovered results they hadn’t thought of but really liked.''} P11 commented, \textit{``Diverse intermediate results are very helpful for inspiring design ideas.''}

\textbf{Kinematic-driven search is beneficial for achieving kinematic harmony across effects.}
We found that introducing kinematic-driven search helps reduce the need for users to manually adjust the spatiotemporal properties of particle effects and facilitates harmony in combined artworks. On one hand, this approach reduces the need for fine-tuning individual effects because \textit{``kinematic constraints can act like templates to align the search results.''} On the other hand, similar kinematic behavior contributes to better kinematic harmony in the composed works. P15 mentioned that under the condition with kinematic-driven search, \textit{``the particle effects look good when placed together, with little need for further manual adjustment.''} Aligning the kinematic behavior during the kinematic-driven search process effectively reduces users’ workload.

\textbf{KinemaFX effectively supports the creation of satisfying effect artworks.}
The particle effects created using KinemaFX received high praise from participants. P1 stated that KinemaFX \textit{``has a low entry barrier,''} as the flexible balance between semantic and kinematic input allows users to \textit{``start creation with only a semantic or kinematic design concept.''} This is useful because users often encounter situations where they have a clear kinematic design but don’t know how to describe it, such as \textit{``wanting to express a ‘tunnel’-like structure but couldn’t find the right term''} (P7). Additionally, simple semantic and graphical inputs can efficiently express core intent while preserving diversity in other features. P14 pointed out that \textit{``simple input can filter out many irrelevant results,''} and during the iteration process, they could \textit{``explore many satisfactory results.''} This approach helps ensure that final works align with users’ core intentions while maintaining diversity in other aspects.

\textbf{KinemaFX reduces the mental effort of creation.}
We found that creating effect artworks using KinemaFX involved less mental effort compared to other conditions. This is partly because graphical input reduces reliance on semantic input, while \textit{``constantly coming up with proper semantics to express intent is exhausting''} (P10). Additionally, the iterative process driven by implicit preference guidance relieves users from the burden of actively thinking through and articulating exploration directions. P3 noted: \textit{``I don’t need to express it myself — I can determine a new direction based on the search results.''} This exploration mode makes the creation process more relaxed.

\section{Discussion}
Based on the development and evaluation process, we primarily discussed the following implications and limitations of our work.

\subsection{Implications}

\textbf{Enhancing the Flexibility of Kinematic Intent Expression.}
Considering that the target users of our work may not possess professional art skills or an understanding of the kinematic behaviors of particle effects, the graphical input used in KinemaFX is simple. The trade-off for lowering the entry barrier for graphical input is that it limits the flexibility of expressing kinematic intent. Enhancing this flexibility could be beneficial when catering to different user groups. For example, in our formative study, we found that some creators are accustomed to using sketches to express kinematic information. The combination of sketches with computing methods for human-computer collaborative creation has already been proven effective~\cite{guay2015space, choi2016sketchimo, yu2023videodoodles}, providing higher flexibility in input for particle effect creation. Additionally, controlling particle effects in VR or AR environments holds promise, as embodied interaction in 3D spaces can lead to more intuitive and natural experiences. Similar methods have already been used in animation generation~\cite{arora2019magicalhands, li2024anicraft, yuan2025alice} and pose control~\cite{ye2020aranimator, zhou2024timetunnel}. Moreover, KinemaFX requires users to actively construct graphical input, while P2 pointed out that users might sometimes \textit{``wish for recommended kinematic behaviors.''} This inspires us to simplify input for users lacking design experience through predefined kinematic templates or intelligent recommendations. For more experienced users, the system can open more precise control features. This flexible, layered interaction strategy can enhance kinematic intent expression for users of different skill levels while maintaining ease of use.

\textbf{Exploring the Transferability of Particle Effects.}
In our user study, we have identified the limitations of using semantics to describe particle effects. On the other hand, using semantics to represent particle effects may limit their transferability. For example, a particle effect that continuously shoots arrows within a range: shooting upwards may represent a buff, shooting downwards may represent a debuff, shooting horizontally forward may represent an attack, and shooting horizontally backward may represent acceleration. This leads us to the idea that the same particle effect, through simple kinematic transformations such as translation, rotation, and scaling, can be adapted to different application needs, even though these four meanings are significantly different from a semantic perspective. KinemaFX considers kinematic transformations such as translation, rotation, scaling, speed control, and color adjustments to support the transferability of particle effects. This method is effective, but it can be further enhanced. Generative AI has been widely used to support the creation of stylized works~\cite{xiao2024typedance, zeng2024intenttuner}. By modeling the appearance and behavior of particle effects in more granular detail and utilizing data-driven methods with generative AI, particle effects can intelligently adapt to different usage scenarios. Furthermore, to address the limitations of semantics in particle effect transfer, it is possible to construct cross-semantic mappings by analyzing the kinematic features of particle effects under different semantic labels~\cite{wang2015deep}. \rv{This transferability allows particle effects to fully realize their polymorphic potential, enabling the creation of a broader range of diverse works from a limited effect database.}

\textbf{Explicit and Implicit Preference-Guided Exploration.}
Our work aims to reduce the effort required for non-expert users to actively express preferences during the creation process by adopting a framework for implicit preference-guided exploration. Implicit preference-guided exploration naturally supports heuristic creation when user intent is vague. However, when users have very clear intentions, what they need is to quickly narrow down to a specific area of the design space through explicit and controllable operations. Users often \textit{``have vague intentions in the early stages of design, which gradually become more defined as exploration progresses''} (P16). This inspires us to combine explicit and implicit preference-guided exploration appropriately to provide different support based on the user’s needs. Previous works~\cite{koch2020imagesense, hoi2024creativeconnect, yuan2025personalized} have supported users in simultaneously controlling the creation process through implicit preferences and explicit intentions, such as expressing implicit preferences through example images and explicitly selecting semantic tags recommended by the system. This collaborative mechanism between explicit and implicit guidance can better adapt to the evolving goals and cognitive states of users during the creation process.

\rv{
\subsection{Generalizability}

While KinemaFX targets the creation of effect artworks, our work can generalize to broader domains. 
At the core of this generalizability is the principle of conveying complex kinematic information accurately and intuitively by representing multiple motions as transformations of a composite shape. This abstraction aligns with how designers often perceive and reason about complex motion.

Our conceptual model applies to other design domains involving intricate kinematic behavior, such as character animation (e.g., for dance or combat) and embodied interaction design. For instance, designing combat animations involves the 3D spatiotemporal coordination of multiple skeletal joints and requires fine-grained control. Similarly, in gesture design, although motions involve multiple joint positions, designers typically conceptualize them based on the overall configuration and transformation of the hand.

Beyond design tasks, particle effects have applications that extend past aesthetics. Kinematic-driven effects are also used in scientific visualization, such as representing large-scale particle movements or simplified fluid dynamics. These use cases share the same structural properties: homogeneous multi-object dynamics and a need for intuitive, expressive motion control.

Our conceptual model facilitates such generalization by decomposing motion design into a semantic component for expressing intent and a kinematic component for capturing spatial and temporal structure. Built on this foundation, our structured representation and distance metrics support a preference-guided, iterative workflow. With appropriate adaptation, this framework can extend to other domains where complex motion representation and control are critical.
}

\subsection{Limitations and Future Work}

Experimental results demonstrate that KinemaFX effectively supports particle effect exploration and customization. However, during the system implementation and evaluation processes, we identified several limitations. We hope to highlight these limitations to inform potential future work.

\textbf{Guidance for effect design.} 
KinemaFX focuses on user preference-guided exploration, and therefore does not provide additional guidance on how to design a complete effect artwork. Potential guidance could include suggesting the components of an effect composition or recommending spatial-temporal layouts for combining multiple particle effects.

\textbf{Better support for particle effect composition.} 
KinemaFX provides basic control over the spatial and temporal attributes of particle effects to support harmonious combination. However, the creation of more complex effects may rely on more fine-grained interaction mechanisms during the composition process, as well as consideration of hierarchical relationships and logical dependencies among particle effects.

\rv{
\textbf{Further evaluation.}
Results and feedback from the user study show that relying solely on LLMs for semantic processing limits effective expression of kinematic information in human-AI collaboration. However, we did not independently evaluate LLM performance or compare different LLM models.
Regarding computational performance, structured representations and rule-based metrics enable much faster processing than LLM-based reasoning. Kinematic extraction and semantic data processing can be done offline, so the main real-time bottleneck is the LLM response time for user semantic input and preferences. We plan to evaluate these performance aspects systematically in future work.
}
\section{Conclusion}

We presented KinemaFX, a kinematic-driven interactive system to assist non-expert users in constructing customized particle effect artworks. KinemaFX combines semantic and kinematic information to support implicit preference-guided exploration of particle effects. We also developed a kinematic-driven method to support more efficient and intent-aligned particle effect search within KinemaFX. We validated the effectiveness of KinemaFX through a user study involving 16 participants. We also discuss the scalability and limitations of KinemaFX, hoping our findings will inspire future research in creative support tools.

\begin{acks}
The authors would like to thank the reviewers for their constructive feedback. This work is supported by the Natural Science Foundation of China (NSFC No.62472099) and Ji Hua Laboratory S\&T Program (X250881UG250).

\end{acks}


\bibliographystyle{ACM-Reference-Format}
\bibliography{ref}

\end{document}